# Aligning load flexibility with emissions reduction: empirical insights from a multi-site study of cryptocurrency data centers

Veronica M. Paez[a]*, Neda Mohammadi[a], John E. Taylor[a]

[a]School of Civil and Environmental Engineering, Georgia Institute of Technology, Atlanta, GA[ib]

* Corresponding author. School of Civil and Environmental Engineering, Georgia Institute of Technology, North Avenue, Atlanta, GA 30332.

E-mail addresses: mpz@gatech.edu (V. M. Paez), nedam@gatech.edu (N. Mohammadi), jet@gech.edu (J. E. Taylor)

## Abstract

The power sector is responsible for 32 percent of global greenhouse gas emissions. Data centers and cryptocurrencies use significant amounts of electricity and contribute to these emissions. Demand-side flexibility of data centers is one possible approach for reducing greenhouse gas emissions from these industries. To explore this, we use novel data collected from the Bitcoin mining industry to investigate the impact of load flexibility on power system decarbonization. Employing engineered metrics to explore curtailment dynamics and emissions alignment, we provide the first empirical analysis of cryptocurrency data centers' capability for reducing greenhouse gas emissions in response to real-time grid signals. Our results highlight the importance of strategically aligning operational behaviors with emissions signals to maximize avoided emissions. These findings offer insights for policymakers and industry stakeholders to enhance load flexibility and meet climate goals in these otherwise energy intensive data centers.

Word count: 7772

## 1. Introduction

Climate change, exacerbated by fossil fuel combustion releasing greenhouse gases, is warming the atmosphere to unprecedented levels [1]. The power sector significantly contributes to this warming, with 32% of global emissions resulting from electricity generation. Data centers (including data-transmission networks), consume an estimated 2.4-3.3% of global electricity. Additionally, cryptocurrencies are known to be major power consumers and are often criticized as rigid loads that exacerbate grid emissions [2]. Bitcoin alone uses an estimated 0.44% of global electricity [3], [4], [5].

As one step toward mitigating the worst outcomes of climate change, countries aim to decarbonize the electrical grid, transitioning to renewable energy sources [6], [7]. To do this, electrical grids must undergo a major transition from conventional fossil-fuel generators to a supply side dominated by wind, solar, and batteries. Typically, power production can ramp up and down to meet changes in demand (i.e., power consumption from business and residential buildings). However, as more variable renewable energy provides power to electrical grids, misalignment between supply and demand becomes more common and a serious threat to power system reliability. Research shows that demand-side flexibility—a load's ability to change its power consumption in response to electricity prices or grid signals—provides a cost-effective approach to integrating renewable energy. This load flexibility improves alignment of consumer demand with energy supply, enabling decarbonization goals to be achieved without compromising grid reliability [8].

Data centers, with their significant power draw, are strong candidates for load flexibility [9]. A 30MW data center could provide the same services as an optimized 7MWh large-scale storage installation [9]. However, traditional data centers have high-reliability requirements that can limit this [10]. Data centers, in most cases, are constrained by Service Level Objectives (SLOs) and typical data center load profiles are steady [11]. Additionally, regulatory and economic barriers disincentivize participation [12].

Data center operators often fear that providing flexibility would incur unacceptable risk or waste expensive hardware and harm user experience [13]. In contrast, cryptocurrency data centers (CDCs) reportedly do not have the same reliability concerns and could be modified to provide the necessary flexibility [14]. In practice, data centers adjust their power consumption based on one or both of the following criteria: (i) in response to incentive payments that induce lower electricity use at times of high wholesale market prices or when system reliability is jeopardized or, (ii) in response to shifts in real-time electricity prices [15]. For CDCs, cryptocurrency exchange price may also influence these adjustments, though a detailed price-incentive analysis is beyond the scope of this study.

The high reliability requirements and steady load profiles of traditional data centers leave open questions regarding how alternative models, like CDCs, could effectively provide load flexibility. The recent *2024 United States Data Center Energy Usage Report* from Berkeley Lab called on researchers to identify opportunities for data center load flexibility as a means to reduce their impact on grids and mitigate their carbon footprint [16]. However, most existing literature addressing flexible loads has relied heavily on simulations or theoretical frameworks applied only to traditional data centers. These simulation-based studies often emphasize workload balancing optimization or demand-response incentive design on idealized electrical grids, leaving significant uncertainty regarding real-world effectiveness.

Operational data from CDCs, which could validate actual behavior and potential for decarbonization, remain scarce. Additionally, blockchain energy models assume CDCs are located on carbon intensive energy grids with constant uptimes. We identified only one study that used empirical data, collected from the Texas' Electricity Reliability Council of Texas (ERCOT) [17]. As a result, our understanding of the effects of geographic differences in energy mixes and operational behavior on real world load flexibility is limited. The authors are not aware of any study that empirically investigates this heterogeneity in CDCs. Our study reveals a critical gap between theoretical and empirical behavior.

Bridging this gap, our study uses novel, real-world energy data from 21 North American CDCs and demonstrates that they are capable of load flexibility and can, to varying degrees, mitigate their greenhouse-gas emissions. We first establish that uptime alone explains only a fraction (adjusted-$R^2$ = 0.19) of avoided emissions. This simple regression model was then used to create four performance quadrants to highlight variation across data center operators. Using a facility-specific threshold algorithm, we computed curtailment events and emission related metrics (curtailment magnitude, induced emissions, maximum energy, and LME variability). We then used these metrics in a stepwise regression, which raised the adjusted-$R^2$ to 0.59. This revealed that a majority of avoided emission variance is explainable through the addition of the facility's maximum energy use and the grid's shifting marginal generator stack. Lastly, we provided case studies that build on these results to highlight operational and grid differences. These results demonstrate that voluntary, price-driven curtailment plays a key role in carbon mitigation and point to targeted pricing and placement strategies as levers to maximize emissions reductions.

Our paper follows the ensuing structure. In 2, we examine the infrastructure and energy consumption of traditional data centers and introduce a subtype known as Proof-of-Work (PoW) CDCs. Focusing on CDCs, we further introduce the unique concept of difficulty-driven power demand, and also discuss demand-side flexibility and the importance of hourly emission modeling for studying load flexibility. Our research design, data, and methods are presented in 3. Finally, our findings, contributions to data center energy research and broader impacts are detailed in 4-6.

## 2 Background

### 2.1 Data Centers & Cryptocurrency Mining Data

#### 2.1.1 Data Centers

Data centers house hundreds to tens of thousands of servers, cooling systems, storage drives, and network devices. Their customers expect consistent availability of resources, so data centers also require backup power, such as uninterruptible power supplies (UPS) and diesel generators.

The International Energy Agency (IEA) estimates that traditional data centers plus data-transmission networks consumed 500-700 TWh of power in 2023 [2]. A significant portion of this demand comes from artificial intelligence (AI), such as large language model services like ChatGPT. In 2023, AI processors were estimated to have used about 7-11 TWh of electricity, or roughly 0.04% of global electricity use [2]. Projections suggest a ten-fold increase in energy use from the 2023 estimate by 2026 [5]. These same forecasts project that along the same timeline, data centers' total electricity consumption could rise to more than 1,000 TWh [5]. This is a significant increase after several years of energy consumption plateauing as a result of efficiency gains in storage capacity, cooling, and server virtualization [18]. The IEA forecast includes traditional, AI dedicated, and CDCs [5].

## 2.1.2 The Rise of the Cryptocurrency Data Center

CDCs operate distinctly from traditional data centers, as they generate revenue not from end-user demand, but from earning cryptocurrency rewards through computational power. While traditional data centers monetize workloads and CPU time based on diverse computational needs and end-user demand, CDCs' energy consumption is driven by the demand for computational power to secure the network's transactions and by the difficulty adjustment. Focusing on Bitcoin, a distributed monetary network, reveals how the Proof-of-Work (PoW) security mechanism consumes energy to secure its public transaction ledger [19]. Although only about 5% of cryptocurrencies use PoW, Bitcoin constitutes roughly 64% of the total cryptocurrency market capitalization[20, 21].

In Bitcoin's cryptographic lottery, or "mining," the first computer to find the winning block receives a reward, which halves approximately every four years (since 2024, this reward is 3.125 Bitcoin). The difficulty adjustment regulates how hard or easy it is to find new blocks of transactions. This lottery has become progressively more challenging over time. The combination of increasing difficulty and decreasing block subsidies intensifies competition among computers involved in block finding, pushing the Bitcoin mining market toward near-perfect competition [22]. Consequently, electricity cost has become a primary concern for Bitcoin data centers [23]. Despite efficiency gains in mining machines, Bitcoin's energy demand still increased to an estimated 121.13 TWh in 2023, raising substantial environmental concerns [3, 24]. In theory, these data centers can power down machines when their break-even price drops below electricity costs to stay competitive, and this difficulty-driven price sensitivity provides an incentive that makes Bitcoin data centers suitable for load flexibility and for potentially reducing their emissions.

**2.2 Load Flexibility**

*2.2.1 Demand-side Flexibility*

Variable power sources like solar and wind often cause supply and demand mismatches on the grid, leading to electricity deficits or surpluses. Increased variable renewable energy penetration, particularly in liberalized markets, can paradoxically reduce profitability and increase volatility, but greater demand-side flexibility can mitigate these concerns [25, 26]. Demand-side flexibility involves various strategies for power consumers to manage electricity demand, including load flexibility, energy efficiency, demand response programs, or distributed energy resources. Our focus here is specifically on load flexibility, a subset of demand-side flexibility, which allows a load to alter its power consumption in response to external factors like electricity prices or grid reliability issues, also known as demand response or ancillary services.

For example, traditional data centers can shift workloads in response to grid signals or pricing changes when energy supply cannot meet current consumer demand. Demand response has a decade-long history, with its origins in the late 1970s, a decline in the early 2000s, and a resurgence in the 2010s due to smart grid investments [27]. Demand-side flexibility can reduce the cost of variable renewable energy and serve as a valuable decarbonization tool [8]. One study found that short-term demand response, combining energy efficiency and demand shedding and shifting, offered the greatest benefits by substituting for gas-peaker capacity and promoting renewable energy growth [28]. Another study noted that reducing peak loads and shifting demand to low-carbon emission times could decrease emissions [29]. A review of nine studies across multiple countries indicated that load flexibility accounts for roughly 2%-20% of decarbonization capacity scenarios [30]. Furthermore, a review of data center demand response literature, focusing on medium-sized data centers, suggests that improved market design and economic incentives could enhance data center participation [11].

### *2.2.2 Data Center Load Flexibility Studies*

### *2.2.2.1 Cryptocurrency Data Centers*

Bottom-up data collection, while offering historical accuracy for data centers, has limited availability and labor-intensive processes [31, 32]. These issues lead to simplistic extrapolations and insufficient data in blockchain energy research, particularly for Bitcoin, often ignoring locational variations and lacking retrospective analysis [33, 34]. Such unreliable findings can misinform policymakers and the public, potentially undermining decarbonization or optimal energy use.

Furthermore, CDC studies often overlook load capacity utilization variations [33] and we found no CDC energy studies considering demand-side flexibility's impact on overall energy use or carbon footprint. Only a few studies model Bitcoin data centers for demand response in the ERCOT region. For example, [35] modeled price-responsive Bitcoin miners on a Texas grid, finding that while increased electricity demand from these facilities doubled wind capacity, emissions also rose, though demand response could offset this. Another study [17] used ERCOT data to model Bitcoin data centers' impact on emissions, market prices, and reliability in Texas, revealing significant location-dependent carbon emission variability and that price-responsive flexibility could mitigate adverse impacts. An optimization problem in [36] showed that strategic participation in ERCOT ancillary services could increase a cryptocurrency miner's revenue by up to 20%, suggesting a new revenue stream for these data centers [37, 38]. [37] showed that well-coordinated cryptocurrency mining data centers in Texas can leverage their flexible loads to provide demand response services, though their size and location can drive non-uniform spikes in locational marginal prices. Lastly, [38] demonstrated using real-world ancillary services data that these mining facilities can physically and economically viably participate in fast-response frequency regulation in ERCOT. While these studies are informative, the focus on ERCOT creates uncertainty about CDCs' effectiveness in diverse grid environments, limiting existing models' generalizability.

### 2.2.2.2 Traditional Data Centers

Numerous papers have explored flexible loads in traditional data centers [39]. For example, [40] modeled utilities setting real-time prices based on grid balance needs, incentivizing data center load flexibility through electricity cost changes with per-site and average price caps. CDCs could likely benefit from this, and extending the model to include pure price responsiveness might reveal different levels of energy savings given additional revenue volatility from network competition and cryptocurrency prices. Similarly, [41] used a game theoretical approach to increase data center participation in demand response with real-time pricing, showing that highly flexible data centers could save 18.7% in contract payments and increase utility revenue by up to 80%. CDCs, lacking workload scheduling needs, could be optimal participants in such markets.

[42] simulated pre-cooling in a telecom data center, finding 7.8-8.6% savings in cooling costs and successful load shifting during peak hours. Another paper [43] developed a convex-optimization framework, demonstrating that off-site generation and workload shifting could reduce costs by 35-40% and emissions by 10-15% during peak hours. While these studies show data centers can reduce peak load, empirical evidence on their actual participation in such programs remains limited. [44] modeled spatiotemporally flexible data centers with waste heat reuse, finding improved grid operation efficiency and reduced carbon emissions through coordinated flexibility, though practical operational evidence is lacking.

Conversely, [45] empirically validated data center demand response through coordinated IT shutdowns, job scheduling, temperature adjustments, and geographic load migration, showing up to 31% rapid load flexibility. However, this study focused on only four California data centers and predates CDCs. Other papers present energy or carbon-aware workload shifting algorithms, and while energy-awareness has minimal application for

CDCs due to server operation differences [45] and [46] are recommended surveys on the subject. Carbon-aware scheduling shifts workload based on hourly carbon intensity. [47] incorporates on-site battery storage for excess solar, while [48] used locational marginal emission factors and a hypothetical carbon tax, finding data centers shifting 0.85% of total system load and reducing emissions by 1.82% without market redesign. However, carbon-awareness workload shifts have limitations, including data center capacity, workload tolerance for long-distance moves, short turnaround for most batch jobs, and governmental/privacy restrictions on global migration [49]. Most studies rely on experimental simulations, with a few tested in physical data centers, like [50] or Google's fleet [51], which showed 1-2% power reduction during peak carbon hours. Critically, CDCs do not need to shift workloads for flexibility, reducing the relevance of such algorithms and highlighting a need for alternative approaches that utilize their simpler design.

### 2.2.3 Carbon Emission Factors for Emissions Estimates

Previous studies estimating CDC carbon emissions primarily used network-wide top-down approaches, employing country-level average grid energy mixes and computational power distribution based on IP addresses. In [52], the authors assumed that the mining pool's headquarters were where the mining computers were located and when a pool's location was unknown, they used the global average. Cambridge University's CBECI also uses mining pools to share geographic data from IP addresses [3], and other studies, like [53], have utilized this data. However, using country-level or global average emissions has limitations. These averages do not account for temporal changes in carbon intensity or interregional electricity flows [54]. Studies like [55] rely on low-resolution annual carbon factors, which can misrepresent a data center's true environmental impact, underestimating emissions by up to 22% or overestimating by 33% [54]. [48] also found that using average emissions over marginal emissions increased emissions in their load-shifting model, suggesting hourly emission factors are better for evaluating data center load flexibility and improving policy and operational efficiency.

[17] developed a granular emissions estimate at the grid bus level for CDC carbon emissions to compute emissions for cryptocurrency miners, revealing high heterogeneity in carbon footprints across their synthetic Texas grid model. Most studies have modeled cryptocurrency emissions at the network level or within a single grid, using top-down simulations or lifecycle assessments lacking temporal resolution like in [56]. Extending these approaches to other regions requires complex simulations, making cross-grid comparisons impractical. In contrast, locational marginal emission factors (LMEs) capture the hourly carbon intensity of marginal power plants, offering the granularity needed for real-time emissions quantification for individual facilities.

## 3. Materials and Methods

This study investigates the real-world behavior of load-flexible cryptocurrency mining data centers and their ability to reduce their marginal carbon emissions. We quantify avoided emissions over a three-month period and analyze how operational load reductions align with temporal grid emissions signals.

## 3.1 Research Design and Data

We chose a quantitative and observational framework for our study. We believe this approach sets a foundation for filling the identified gaps in the literature: (i) limited empirical studies exploring load flexibility in energy-intensive operations, (ii) lack of real-world data to validate assumptions, and (iii) minimal research on how load flexibility impacts energy use and emissions in cryptocurrency mining data centers.

For the study, we contacted 38 private and public CDC providers and received data from 12, or roughly 32% of those we contacted. In 2024, The EIA identified 137 facilities in the United States, with 52 of those with known location and capacity data [57]. Ultimately, our study included data from 21 facilities with known location and capacity. Our participant pool included only Bitcoin mining data centers that were located in North America. Each of these facilities contain thousands to tens of thousands of computers. We excluded participants who could not provide historical, continuous, hourly, or higher-resolution energy consumption data. We also excluded those that could not offer more than one month of data. To incentivize data sharing, we agreed to anonymize and aggregate the energy consumption data at the state or province level.

Data were collected on-site at the data centers using industry-standardized equipment. Data were cross-verified using available data from ERCOT's large flexible load taskforce presentations to ensure that the load profiles were what would be expected from these types of facilities [58]. Finally, we filled in missing hours with null values. We acknowledge that discrepancies in equipment performance could affect the results of our study.

We received data from a diverse set of locations. One facility was located in Quebec, Canada, while the remaining were situated across seven states: New York, Georgia, Ohio, Kentucky, North Dakota, North Carolina, and Texas. Thirty-three percent of the facilities were located in Texas. Twenty-four percent of facilities were located in Georgia. The EIA study noted that most of the facility they identified were in Texas, Georgia, and New York [57]. In total, we received data for 22 facilities, but we excluded one facility based on the criteria developed in the research design. For those that sent 15-minute interval data, we averaged across the sets of 4 intervals to get an hourly value.

We identified near-node locational marginal emission factors using the facility's physical, geographical location. We collected nodal LMEs from RESurety and a sub-regional LME set from WattTime for the Canadian site [59], [60]. Marginal emission factors reveal what happens to the emissions rate of the generators that are responding to changes in load on the local grid at a certain time [59]. Practically, LMEs are based on the emissions of the marginal power plant that would generate the next unit of electricity. This means that these factors are in units of tons $CO_2$/MWh. As a result, these factors are inherently tied to the marginal generator's emissions and are thus independent of the load capacity of the energy consumer. These are emission rates, so they scale the emissions linearly.

To maximize the duration of analysis, we chose months that overlapped across the maximum number of facilities for which data were received. The result of this design choice meant that despite some companies sharing data that spanned 12–24 months of continuous hourly energy consumption data, we could only include three months in our study. Our study window spanned July 1-September 31, 2023.

## 3.2 Methods

### 3.2.1 Feature Engineering

Feature engineering is crucial in machine learning and statistical modeling, transforming and selecting input variables for models. For example, authors used feature engineering to analyze daily energy consumption time series for cluster analysis, detecting atypical consumer energy use [61]. Building on this, we created and assessed multiple independent variables from energy consumption and LME time series data. These variables serve as metrics for quantitatively investigating data center energy consumption and its load flexibility. We used uptime as a foundational indicator of facility activity. Uptime signifies the total hours a CDC operated at full load capacity, excluding hours with reduced power due to suspected grid signals or real-time pricing. We also employed an energy use threshold to differentiate actual load flexibility events from noise or underclocking. While uptime offers a useful foundational metric, its limitations in capturing nuanced operational behaviors necessitated testing additional engineered metrics through forward-backward stepwise regression. Consequently, we developed metrics to capture curtailment dynamics and LME alignment.

Curtailment metrics provide a multidimensional view of curtailment behavior, complementing the use of uptime as the baseline predictor. Curtailment dynamics included curtailment magnitude (CM) and curtailment regularity (CR). These metrics provide insight into how a facility adjusts its load in response to external signals.

Curtailment magnitude (Equation 1) is the average percentage reduction in energy consumption relative to the baseline energy consumption during curtailment periods. It uses the lowest 25% quartile to isolate the lowest energy consumption hours to focus on best-performance curtailment.

$$CM = 100\left(1 - \frac{\overline{E}_{lowest\ 25\%\ DR}}{\overline{E}_{non-DR}}\right) \quad (1)$$

Where the numerator is the average energy in the lowest quartile of curtailment hours and the denominator is the daily average during non-curtailment hours.

Curtailment Regularity (Equation 2) quantifies how consistently a facility initiates its curtailment events. It is computed as the standard deviation of all start times divided by 3,600 seconds, yielding the variability in hours.

$$CR = \frac{\sigma_{start\ times}}{3600} \quad (2)$$

Similarly to the utility of the curtailment metrics, we defined maximum energy (ME), emissions ratio (ER), normalized avoided emissions (NAE), and LME variability (LMEV) as additional alignment metrics.

Maximum Energy (Equation 3) represents the highest hourly energy consumption observed at a facility during the three-month window. We used this metric to normalize other metrics and explain scale-related effects.

$$ME = \max_{d \in D} E_{d-max} \quad (3)$$

Where $E_{d\text{-}max}$ is maximum energy found over a 24-hour period.

Normalized Avoided Emissions (tons $CO_2$/MWh, Equation 4) takes the total avoided emissions divided by the facility's maximum energy use. The normalized avoided emissions metric was used for performance comparisons across different facility sizes.

$$NAE = A/E_{max} \quad (4)$$

Where $A$ is the total avoided emissions (tons $CO_2$) and $E_{max}$ is the facility's maximum hourly energy consumption (MWh). This metric enables cross-facility performance comparisons.

Emissions Ratio (Equation 5) is a unitless measurement of avoided emissions divided by induced emissions. We used this metric in our case studies to rank facilities by carbon effectiveness.

$$ER = A/I \quad (5)$$

Where $A$ is avoided emissions and $I$ is induced emissions.

LME variability (Equation 6) represents the standard deviation of the locational marginal emissions for a given facility. It reflects fluctuations and timing of grid emissions.

$$LMEV = \sigma_{LME} \quad (6)$$

Where $\sigma_{LME}$ is the standard deviation of the LME values in the group.

We selected avoided emissions as the dependent variable and defined it as the amount of emissions that could have been produced if the data center had not reduced its power in response to price or grid signals.

Our understanding of avoided emissions is based on the Greenhouse Gas Protocol, a framework commonly used by corporations to report their carbon footprint [62].

Following their recommendations for the consequential approach, we take the emissions in the baseline scenario (no load curtailment) and subtract the emissions in the policy scenario (load curtailment). To compute the avoided emissions, we first time-aligned and merged the energy consumption and LMEs to ensure synchronized records.

As shown in Equation (7), to compute the avoided emissions for a given hour, we subtract the energy used in an hourly interval from the daily average for hours that were flagged as false, and multiply the LME for that hour against the difference of those values.

$$A_{id} = (\bar{H}_{d,false} - e_{id})LMEF_{id} \quad (7)$$

Where $\bar{H}_{d,false}$ is the average non-curtailed energy on day d, $e_{id}$ is the energy used in hour $i$, and $LMEF_{id}$ is the locational marginal emission factor for that hour.

Next, we extracted columns for the hour, day, weekday, and month to facilitate the time-based analysis. We replaced outlier energy values that were 1.5 times higher than the previous day's maximum with the nearest valid (non-outlier) value. This step ensured that extreme or erroneous values did not interfere with curtailment detection.

### 3.2.2 Curtailment Detection

The curtailment detection procedure used in this study is related to earlier works that focused on load forecasting, customer baseline load predicting, and day of demand response event detection [63]. Load forecasting is the prediction of energy use on normal days and the customer baseline load estimates the consumer's energy consumption had the demand response event not occurred. Inspecting the energy consumption data (Figure 1) showed that these data centers generally operate as a constant energy consumer except when curtailing energy consumption. Knowing this allowed us to simplify our approach compared to typical customer baseline load methods. With this in mind, we used the daily maximum energy consumption as a reasonable approximation for the typical baseline load when curtailment would not have occurred on that day.

Based on industry discussion, we also knew that these types of data centers sometimes implement underclocking. To account for this as well as noise, a threshold was necessary to determine true curtailment events. Using the Kneedle method [64], we performed a sensitivity test to determine the facility-derived threshold. The threshold was defined as a percentage of the daily maximum energy consumption. We normalized the mean $DR_{percent}$ at each threshold across each of the 21 facilities and found the knee point in the system. The knee point across each facility ranged from 75% to 95%, with the mean at 88%, the median at 90%, and a standard deviation of 0.05%. We interpreted these values to mean that anything above these thresholds reflected noise or underclocking and were ignored. Energy values below this threshold were flagged as curtailment using the $DR_{flag}$ metric. The $DR_{flag}$, when set as True, identifies an hour as active curtailment. Curtailment start and end times were identified by transitions in the $DR_{flag}$. Events were marked as active if power consumption remained below the threshold (Algorithm 1, Figure 1).

**Algorithm 1** Curtailment Detection Algorithm
---
**Require:** Energy consumption data $E(t)$, daily maximum energy threshold MaxEnergy, underclocking threshold UnderclockThresh
**Ensure:** Curtailment periods identified with flags $DR_{\text{flag}}$, $DR_{\text{active}}$
1: **for** each day $d$ in dataset **do**
2:    Compute $\text{MaxEnergy}(d) \leftarrow \max(E(t))$ for all $t$ in day $d$
3:    **for** each time interval $t$ in day $d$ **do**
4:      **if** $E(t) < 0.7 \times \text{MaxEnergy}(d)$ **then**
5:        Set $DR_{\text{flag}}(t) \leftarrow \text{True}$
6:      **else**
7:        Set $DR_{\text{flag}}(t) \leftarrow \text{False}$
8:      **end if**
9:    **end for**
10: **end for**
11: **for** each time interval $t$ in dataset **do**
12:    **if** $DR_{\text{flag}}(t) = \text{True}$ and $DR_{\text{flag}}(t-1) = \text{False}$ **then**
13:      Mark $t$ as curtailment_start
14:    **else if** $DR_{\text{flag}}(t) = \text{True}$ and $DR_{\text{flag}}(t+1) = \text{False}$ **then**
15:      Mark $t$ as curtailment_end
16:    **end if**
17: **end for**
18: **for** each time interval $t$ in dataset **do**
19:    **if** $t$ is between curtailment_start and curtailment_end **then**
20:      Set $DR_{\text{active}}(t) \leftarrow \text{True}$
21:    **else**
22:      Set $DR_{\text{active}}(t) \leftarrow \text{False}$
23:    **end if**
24: **end for**

*Algorithm 1. We developed an algorithm to detect curtailment windows based on a minimum curtailment threshold and flags. This approach allowed us to capture the start and end of curtailment.*

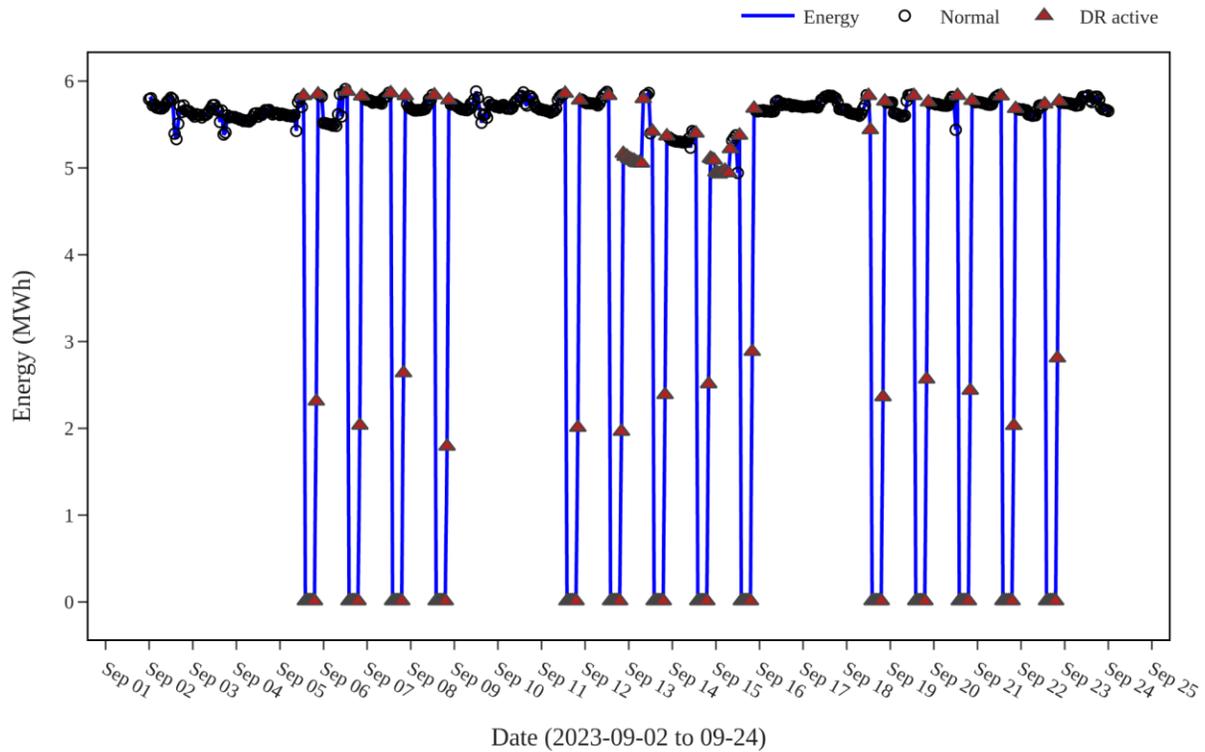

*Figure 1. Energy consumption from September 2 to September 24, 2023, showing normal operation points and DR$_{active}$ points highlighted.*

When the energy consumption for a given hour within a period of 24 hours falls below the threshold, we flag this hour as true for active load flexibility. When it did not fall below the threshold, we flagged it as false. Next, we found the average energy used for the hours that were flagged as false. Using these results, we calculated our metrics from section 3.2.2.

### 3.2.3 Regression Analysis

Given the nature of computing avoided emissions, we know there exists a linear relationship between energy saved (the energy that would have been used had the data center not curtailed) and the emissions avoided. We also know that a change in emissions and a change in load are highly linear [64]. Since energy is power times hours of operation, we can reasonably expect that a regression analysis would be an appropriate starting point for our analysis. Using this understanding, we employed a forward-backward selection, stepwise regression modeling approach to assess the relationship between uptime, curtailment dynamics, LME variability and avoided emissions. The baseline model included uptime as the sole explanatory variable. Additional variables, including curtailment dynamics (regularity, magnitude) and LME variability, were introduced iteratively. Variables were retained only if they improved the model's explanatory power and were statistically significant ($p \leq 0.05$). This approach allowed us to identify the key operational and grid-aligned factors influencing facility performance while ensuring only meaningful predictors were included without the risk of overfitting. For the regression modeling, we used ordinary least squares (OLS) for estimating the parameters of univariate and multivariate linear regression models.

### 3.2.4 Performance Categories and Case Studies

Building on the regression analysis, performance categories were developed to better understand and compare how different facilities translated load flexibility into emissions outcomes. To capture these patterns, we categorized facilities into four groups based on whether they had high or low uptime and high or low avoided emissions. The data center industry generally requires high uptime–better known as availability. Within the industry, this is discussed as the number of nines, often ranging from 99.9% ("the three nines") to 99.9999% ("the six nines"). However, this definition of uptime or availability is subject to debate and the Uptime Institute has remarked that the meaning is not scientifically rigorous [66]. Additionally, our data spans three months, rather than the full year, so using the industry standard for determining high uptime may not be accurate on a shorter timeframe. Therefore, to provide a statistically rigorous cutoff, we used the 75th-percentile quartiles as thresholds. This created a diagnostic framework to visually and analytically identify which facilities were effectively using their flexibility to reduce emissions—and which were not. The performance categories serve both as a comparative tool and as a bridge between statistical modeling and operational insight, offering a way to assess emissions effectiveness across heterogeneous data center behaviors.

Finally, we developed case studies that ground the statistical findings in real-world operational behavior. We closely examined individual facilities to illustrate how differences in their emissions outcomes emerge not simply from facility size or uptime, but from distinct curtailment patterns, grid conditions, and alignment with marginal emissions.

## 4. Results

### 4.1 Comparative Analysis of Regional Emission Profiles

The geographic distribution of the data centers in our study are across the central and eastern portion of North America (Figure 2). When normalized by maximum energy consumption over the three-month study period (expressed as normalized avoided emissions; tons $CO_2$/MWh), the facilities in Texas exhibited similar emission profiles. Facilities in Georgia had the widest range of normalized avoided emissions, and include a facility that avoided the most emissions per megawatt-hour in our study (Facility 18). We pay particular attention to this facility in our case study section. Facilities in Kentucky, North Carolina, and Quebec all showed minimal emission savings, suggesting that these facilities are among the least flexible in our study. We explore this further with our performance categories. The facilities—in North Dakota, Ohio, and New York—showed similar normalized avoided emissions profiles to Texas.

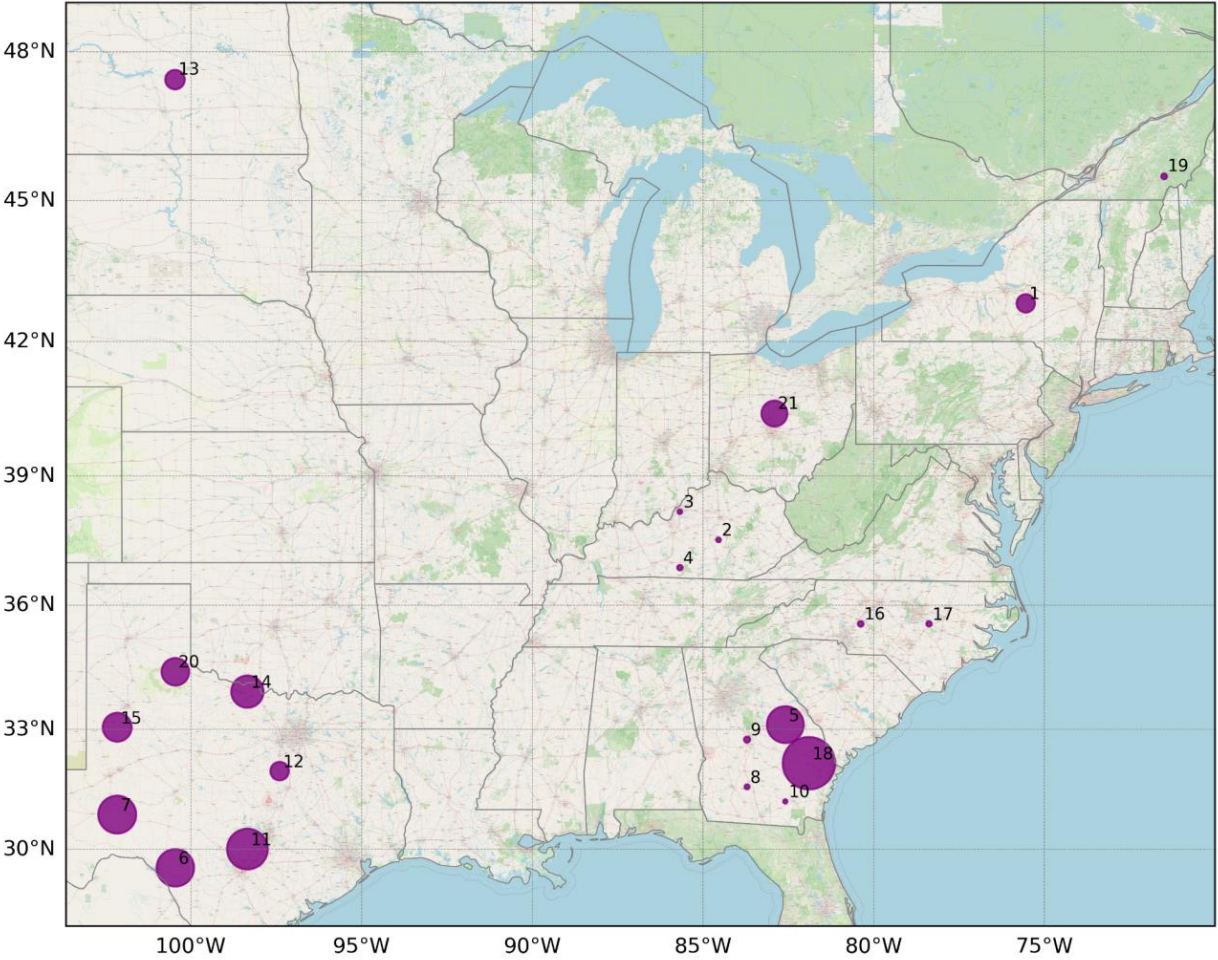

*Figure 2. Geographic distribution of 21 Bitcoin data centers across North America. Marker size represents normalized avoided emissions (tons $CO_2$/MWh). Avoided emissions were normalized by facility maximum energy use for comparison purposes. Marker locations do not represent actual locations beyond the state or province level.*

## 4.2 Regression Analysis and Performance Categories

The results of our baseline regression reveal that uptime explained only about 19% of the total variance (adjusted $R^2$ = 0.19; $p < 0.05$), showing that for CDCs in our study, simply powering down is not a strong predictor for the total absolute emissions avoided from load flexibility. Uptime had a negative effect on avoided emissions ($\beta$ = –197.93, SE = 69.07, t = –2.86, p = 0.01), meaning that the more hours the facility was consuming energy, the lower its emissions saving.

We further explored this relationship using performance categories and found that the facilities fell within three types out of the performance categories (Figure 2). The High Uptime & Low Avoided Emissions group included 6 facilities (Facility 3, 4, 8, 10, 17, 19). The Low Uptime & Low Avoided Emissions group included 9 facilities (Facility 1, 2, 3, 9, 12, 15, 16, 18, 20) and the Low Uptime & High Avoided Emissions group included 6 facilities (Facility 6, 7, 11, 13, 14, 21). Theoretically, a fourth category, High Uptime & High Avoided Emissions, was possible, but none of the facilities aligned to this group.

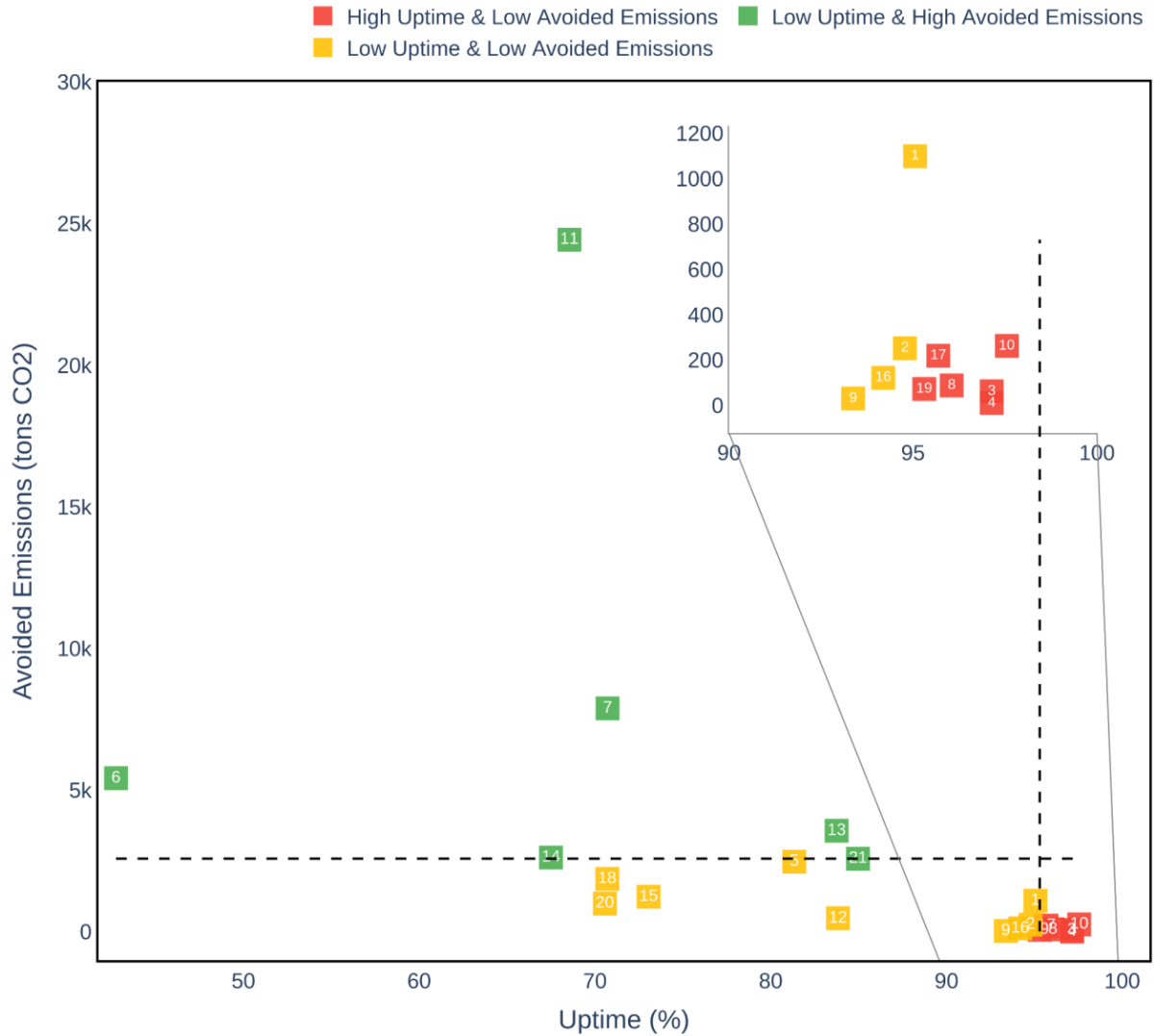

*Figure 3. The baseline regression derived performance categories. Categories were defined using* uptime's distribution mean and *avoided emissions distribution mean.*

Building on the results from the baseline model and the performance categories, we turned toward the forward-backward stepwise regression to further investigate the drivers of these differences. In this regression, we tested the engineered metrics and found two statistically significant features ($p < 0.05$) in addition to uptime. Among the predictors, maximum energy had a significant positive association with avoided emissions ($\beta = 73.24$,

SE = 20.09, t = 3.65, p = 0.002). LME variability also had a significant positive effect (β = 43,250, SE = 17,000, t = 2.54, p = 0.021). Uptime was again negatively associated (β = –213.32, SE = 53.41, t = –3.99, p = 0.001). The combined explained variance (adjusted $R^2$) from uptime, LME variability, and maximum energy increased significantly from 0.19 to 0.59. This is a 210% improvement in the regression model's explanatory power. The correlation between maximum energy and LME variability was 0.067, demonstrating that these features capture essentially independent dimensions of performance. To understand how these performance drivers manifest in practical operational differences, we provide the following three case studies. In them, we continue to use uptime, LME variability, and maximum energy as key explanatory drivers for operational and emissions differences, but also utilize curtailment regularity, curtailment magnitude, emissions ratio, and normalized avoided emissions to describe operational and emissions-reduction patterns.

### 4.3 Case Studies

Given the broad range of operational behaviors, we now present a detailed investigation of individual facilities to better understand the different performance behaviors. We examine facilities from each of the performance categories to better understand their characteristics. These include Facility 11 (Low Uptime & High Avoided Emissions), Facility 10 (High Uptime & Low Avoided Emissions), and Facility 18 (Low Uptime & Low Avoided Emissions).  We plotted our regressors, LME variability and maximum energy, along with descriptive metrics normalized avoided emissions, and emissions ratio for all facilities in Figure 4, to clearly show the comparative effectiveness of emission reductions. Table 1 shows the results of all metrics and regressors for the 21 CDCs.

Comparatively, the LME variability across facilities ranged from 0.10-0.25 tons $CO_2$/MWh (cross-facility σ = 0.046 tons $CO_2$/MWh). For context, PJM reported in 2022 an on-peak LME variability of 0.115 tons $CO_2$/MWh and an off-peak LME variability of 0.122 tons $CO_2$/MWh [67]. In the same year, their marginal generator mix was 87.67% fossil fuel based, mainly gas. Even so, variability arises from switching from gas to coal or other resources from time-to-time. For the facilities in our study, lower LME variability (≈ 0.10 tons $CO_2$/MWh) indicates the marginal generator stack minimally shifts between units with very different emission rates (e.g. mostly similar gas-fired units). Conversely, higher variability (≈ 0.25 tons $CO_2$/MWh) reflects a stack that cycles through a wider range of emission rates. This could be due to cycling between gas and coal, or occasional entry of zero-emitting generators like wind and solar.

While higher LME variability does not indicate more clean energy is meeting changes in load, it does show noticeable hourly shifts in the marginal generator's emissions that, when coupled with curtailment alignment, affect the total avoided emissions. Maximum energy spreads across a wider range, from 3.34-128 MWh (cross-facility σ = 38.42 MWh) among the data centers in our study. Since our time series readings are hourly, we can use maximum energy as a proxy for total data center load capacity and conclude that our facilities span small to hyperscale sized loads.

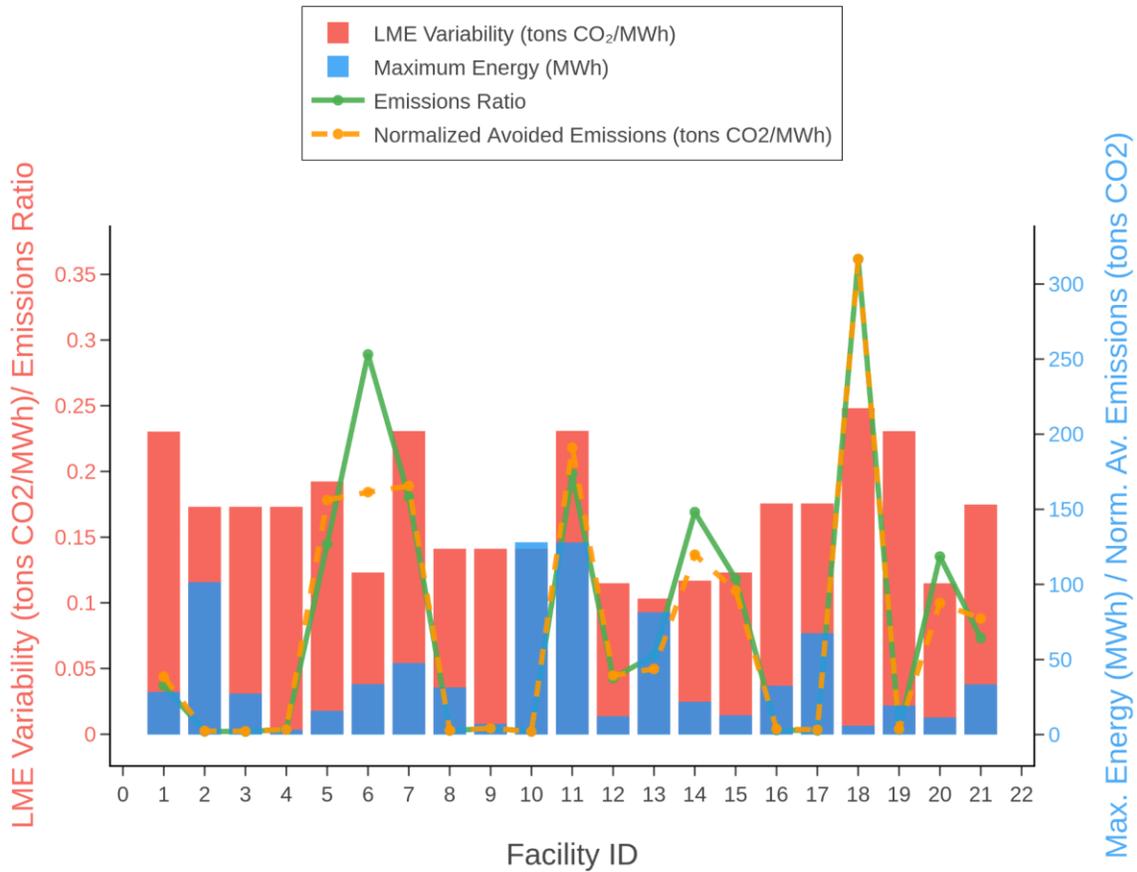

*Figure 4. We see distinct operational profiles across the 21 facilities in our study. The combination of grid LME variability, the facility's maximum energy use, emissions ratio, and normalized avoided emissions show a heterogenous mix of operational choices that encourage a more detailed study to best understand how to utilize these types of data centers for load flexibility. This figure shows Facilities 6, 11, and 18 as the most effective at emissions reduction on a per megawatt-hour basis.*

| Facility ID | Uptime (%) | Avoided Emissions (tons CO2) | Induced Emissions (tons CO2) | Maximum Energy (MWh) | Normalized Avoided Emissions (tons CO2/MWh) |
|---|---|---|---|---|---|
| 10 | 97.55 | 262.17 | 110471.36 | 128.03 | 2.05 |

| 11 | 68.54 | 24433.34 | 123310.33 | 127.9 | 191.03 |
| 18 | 70.7 | 1870.48 | 5172.03 | 5.91 | 316.6 |
| Facility ID | LME Variability (tons CO2/MWh) | Curtailment Regularity (hours) | Curtailment Magnitude (%) | Emissions Ratio | Threshold |
| --- | --- | --- | --- | --- | --- |
| 10 | 0.14 | 7.71 | 24.65 | 0.0024 | 0.9 |
| 11 | 0.23 | 4.2 | 98.57 | 0.2 | 0.9 |
| 18 | 0.25 | 0.42 | 99.83 | 0.36 | 0.85 |

*Table 1*. The upper sub-table shows uptime, total avoided and induced $CO_2$ emissions, maximum energy and normalized avoided emissions for facilities 10, 11 and 18. While the lower sub-table shows locational marginal emissions variability, curtailment regularity, curtailment magnitude, emissions ratio and curtailment threshold for the same set of facilities.

### Facility 10

Facility 10, located in Georgia, represents a High Uptime & Low Avoided Emissions facility (Figure 5). It avoided 262.17 tons $CO_2$ with an uptime of 97.55%, LME variability of 0.14 tons $CO_2$/MWh, and maximum energy of 128.03 MWh, making it one of the largest data centers in our study. Taking a more detailed look at the emissions patterns, when looking at the normalized avoided emissions, the facility avoided 2.05 tons $CO_2$/MWh and its emissions ratio (avoided to induced emissions) was 0.0024. This facility was minimally flexible during the study period, with only one curtailed event detected that lasted 19 hours. Given the steady energy use, it is possible that this event may not be related to grid signals because it occurred during a transition from a lower period of steady energy use to a higher period of steady energy use (Figure 5a). Additionally, a steady load profile suggests that this facility is likely on a fixed price contract, which suggests that it is not likely to be curtailing in response to price signals.

While maximum energy explains part of the avoided emissions, characterizing the way the energy is reduced helps us contextualize the regressor within the facility's operations. The curtailment magnitude was 24.62%, indicating that the single curtailment event was shallow. Not only was facility 10's load flexibility shallow, it was also irregular (CR = 7.71 hours), further emphasizing that this facility is primarily inflexible and that curtailment is rare. Even when there was a reduction in energy use, the low LME variability, shallow curtailment, and poor alignment between energy use and LME over time (Pearson r = -0.03) are all suggestive of a large CDC with an operational strategy that maximizes uptime over emissions reduction or grid balancing.

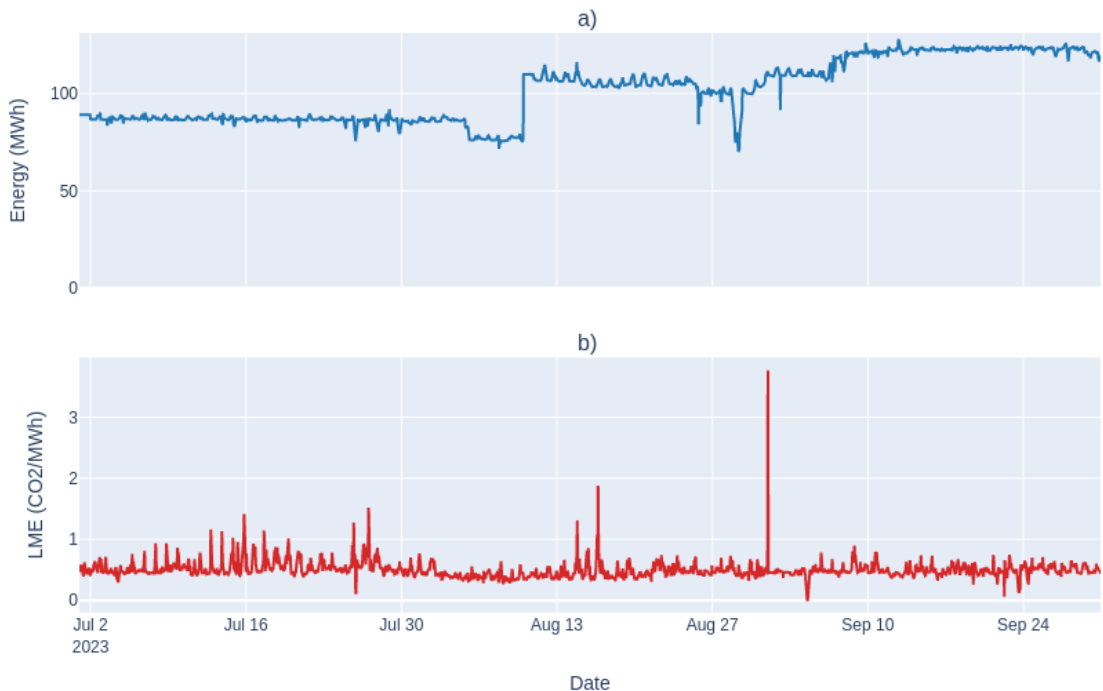

*Figure 5. Facility 10's (a) load profile is relatively steady and shows an increase in energy use over the three-month study period rather than extensive load flexibility. Additionally, there is no overlap between (b) LME behavior and curtailment.*

### Facility 11

Located in Texas, Facility 11 is an example of a CDC in the Low Uptime & High Avoided Emissions group. This facility had an uptime of 68.54%, avoided 24,433.34 tons $CO_2$, had an LME variability of 0.23 tons $CO_2$/MWh, and maximum energy of 127.9 MWh. Compared to Facility 10, Facility 11's maximum energy was nearly identical (127.90 MWh vs. 128.03 MWh). However, as we can see in Figure 6, Facility 11's load profile differs significantly and its LME-to-energy correlation was weakly inversely correlated (Pearson's r = -0.15), suggesting that curtailment behavior is weakly influenced by the marginal generator's emissions.

Looking closer at the emissions patterns, the normalized avoided emissions were 191.04 tons $CO_2$/MWh and the emissions ratio was 0.2. Facility 11's curtailment regularity was 4.42 hours, indicating some irregularity in timing of curtailment events. In other words, while we see in Figure 6 substantial flexibility, we are not seeing a strong pattern for when curtailments occur over the hours. The curtailment magnitude was 98.57%, indicating deep curtailment or near complete reduction of energy use. Given the combination of higher LME variability, its overall capacity, many deep curtailments, and poor alignment with LMEs, we can conclude that this facility is likely price responsive but that the price signals are not well-aligned with the power generator's carbon intensity.

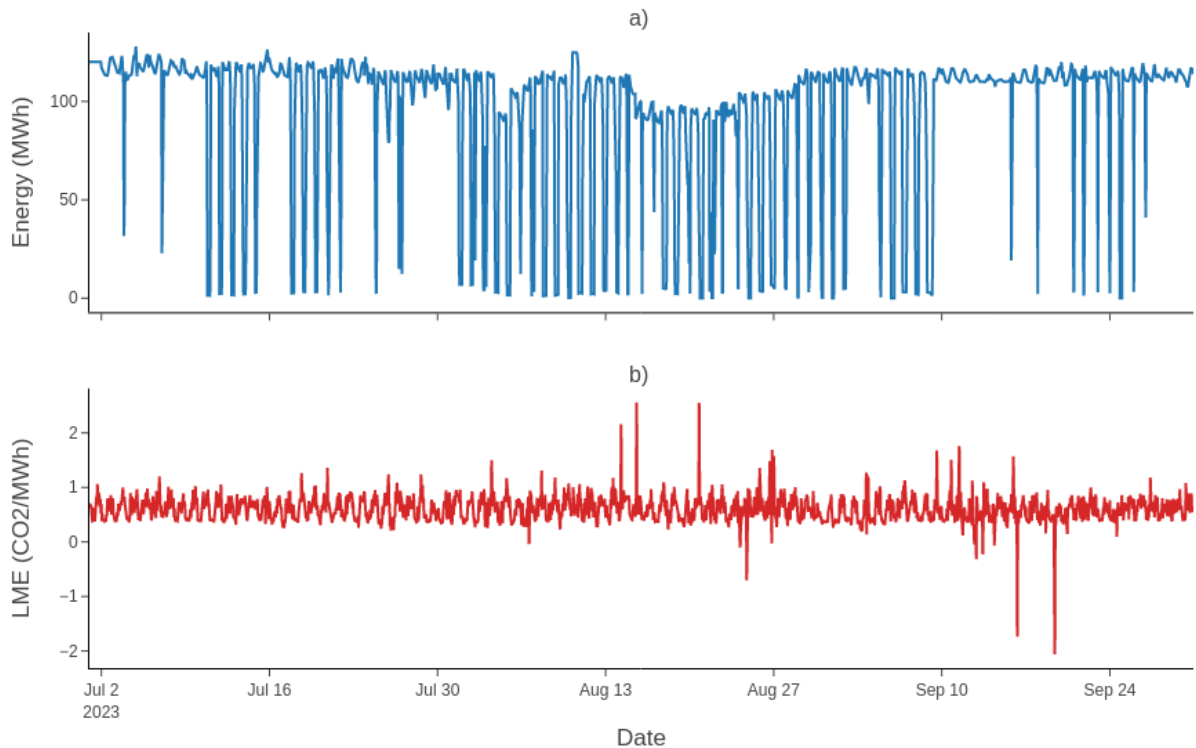

*Figure 6. Facility 11 shows extensive and (a) deep curtailment behavior and relatively (b) high LME variability. Noticeably, there are some negative LME hours, which could either be data artifacts or signs of renewable curtailment.*

### Facility 18

Facility 18 is a member of the Low Uptime & Low Avoided Emissions group and is located in Georgia. It avoided 1870.48 tons CO2, had an uptime of 70.7%, with LME variability of 0.25, and maximum energy of 5.91 MWh. When examining the emissions patterns, we found that this data center's normalized avoided emissions were 316.6 tons $CO_2$/MWh and its emission ratio was 0.36. Compared to other facilities in our study, on a per-MWh basis, this facility avoided the most emissions. The curtailment patterns explain this result. Facility 18's curtailment magnitude was 99.83% and its curtailment regularity was 0.42 hours, suggesting a deep and predictable hourly curtailment pattern. Upon closer inspection of the load profile, we see that the facility usually curtailed between 2pm-9pm daily (Figure 7). Georgia Power's summer on-peak hours are from 2pm-7pm, so this facility's price contract likely requires it to reduce its consumption during on-peak hours. We found that their LME-energy correlation was moderate and inverse (Pearson's r = -0.45), suggesting that curtailments are aligning with higher LME hours. While in absolute terms, this facility avoided fewer emissions than larger facilities in our study, its operations patterns reveal a facility that is located in a region where LMEs and electricity prices are better aligned than in other locations in our study. This alignment, along with the higher LME variability and predictable curtailment, likely explain its per-megawatt-hour effectiveness at emissions reduction.

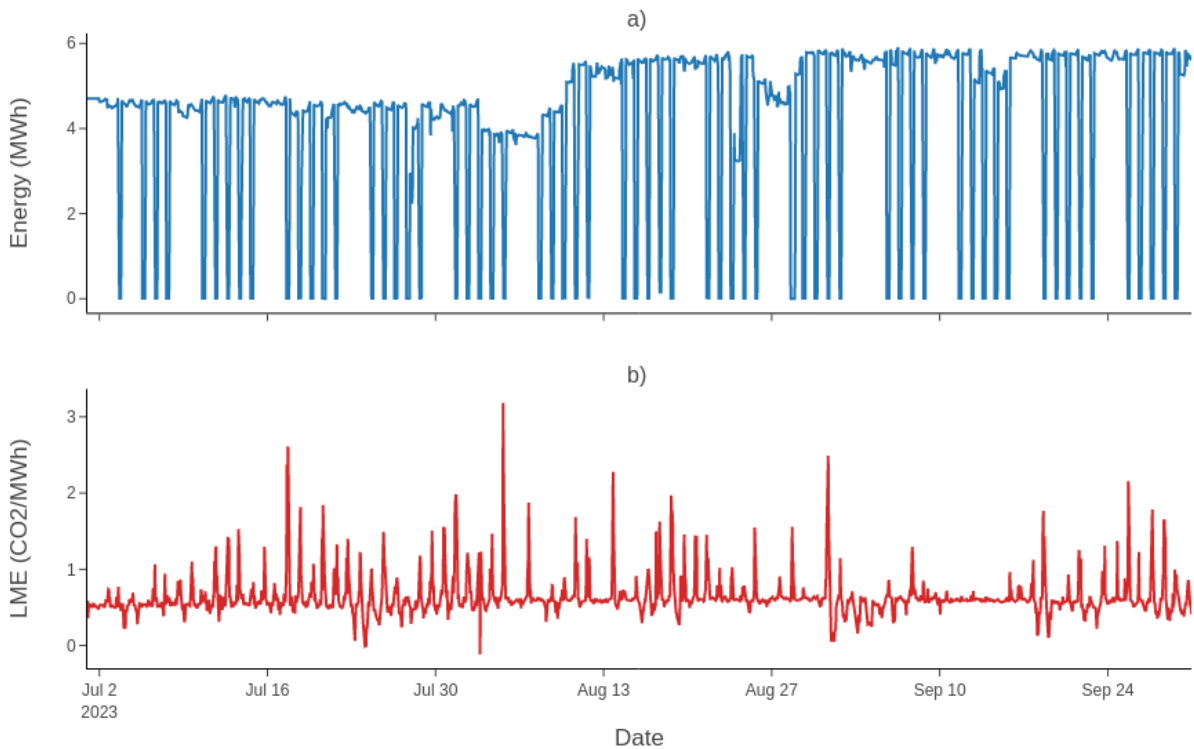

*Figure 7.* Facility 18 shows a (a) highly predictive and steady curtailment profile. Additionally, the is (b) high LME variability, with carbon-intensive marginal generators during certain hours. As we found in this facility's case study, there appears to be alignment between the curtailment behavior, peak and off-peak hours, and low and high LME values.

## 5. Discussion

This study presents the first large-scale empirical assessment of real-time load flexibility in cryptocurrency mining data centers. It builds on previous work [50, 51] that empirically demonstrated load flexibility in traditional data centers using controlled interventions. Unlike those studies, we analyzed facility energy consumption data to investigate their load flexibility behavior. While earlier research on cryptocurrency data centers focused solely on Texas [17], our study expands the geographic scope across multiple states and provinces.

Building on the findings of [54], where the authors showed that average emission factors can significantly misrepresent a data center's carbon footprint, we use near-node marginal emission factors to improve accuracy, except in the single Canadian case where we used a sub-regional set. This method addresses the limitations of earlier approaches that relied on national or global averages [52]. Rather than limiting our analysis to emissions accounting, we used these marginal factors to investigate operational behavior, a topic which remains largely unexamined in prior literature.

Our regression analysis extends previous work [68] that used linear modeling to explore variation in avoided emissions across nine cryptocurrency facilities. In our analysis, we found that uptime alone explained only 19% of the variance in avoided emissions, indicating that curtailment frequency does not directly correspond to emissions outcomes. Incorporating locational marginal emissions (LME) variability and maximum energy increased the adjusted $R^2$ to 0.59. This aligns with theoretical expectations: in a linear relationship, avoided emissions should scale with the product of energy curtailed and the relevant LME. However, because LME values fluctuate by hour and region, uptime serves as a reliable predictor only when facilities are located in similar grid contexts and curtail during similar emissions conditions. Including LME variability resolved this discrepancy. Differences in facility size also justify including maximum energy as an explanatory variable.

Our model confirms substantial variation in siting and operational behavior across CDCs. To reduce emissions effectively, operators should consider both regional marginal emissions and facility scale. When we normalized avoided emissions by maximum energy, deep curtailment and alignment with high-emission hours emerge as the most effective strategies, independent of facility size. These findings support the strategic siting of CDCs in locations where market incentives align with emission signals.

The performance categories derived from our regression highlight distinct operational profiles. Facilities with high uptimes are unresponsive to hourly emissions regardless of location, and low uptime alone did not guarantee high emissions reductions when ignoring facility size. Instead, our results point to an optimal flexibility range, shaped more by regional grid dynamics than by facility capacity.

Our findings contrast with those of [17], which reported no significant emissions reduction from load flexibility due to their sample's baseline price responsiveness and location in low-cost regions. In their model, miners already curtailed during low-price periods, leaving limited additional flexibility when applying their rule to shut down below $40/MWh.

Coordination between market signals and data center operations plays a critical role in emissions outcomes. [48] emphasized the importance of synchronizing facility behavior with LME and LMP data. Our study supports this claim, showing that CDCs can respond effectively to well-aligned signals (e.g. Facility 18). Our results begin to offer operators and policymakers actionable ways to utilize load flexibility in cryptocurrency data centers for improved emission reductions.

Finally, variation in curtailment behavior across facilities reflects underlying financial drivers. Electricity prices, market volatility, block reward schedules, and Bitcoin price cycles influence operational decisions. To support environmentally optimal behavior, future research should incorporate market price data and model facility-level revenue. Policy simulations that combine carbon pricing, real-time demand response incentives, and dynamic electricity pricing could help design effective structures that encourage both emissions mitigation and grid support.

Our study has several limitations. Inconsistencies in third-party metering may contribute to measurement noise. Constraining the study to the summer months introduced some limitations on generalizability across the year, as some facilities may be more or less flexible at different times of year. The performance categories are only reflective of the data centers that participated in our study, and do not present an absolute measure of performance for non-study CDCs. Our curtailment detector identifies drops in energy use over time and avoids noise and underclocking, but with regard to the higher uptime facilities (≥ 95%) that experienced only a few events, some of these could have been operational, so it is possible that their uptimes were slightly underestimated. Lastly, our use of the highest maximum energy value assumes that the energy use is constant except when curtailment events occur. However, some of the data centers in our study were increasing their energy consumption over the three-month timeframe, so it may underestimate the normalized avoided emissions for some of the facilities.

## 6. Conclusion

Our findings have direct implications for policy, operations, and climate mitigation strategies. In 2024, Lawrence Berkeley Lab hosted a workshop on behalf of the Department of Energy regarding data center load flexibility, where they identified the immediate need for utilizing data centers for demand response [12]. Yet, the discussion overlooked CDCs as a readily available solution. This, along with the results of the 2024 EIA report [57] which estimated that CDCs could be using as much as 2.3% of the US electricity, with over 10 GW of power demand, show that our study fills an important gap in understanding. CDCs—with a focus on Bitcoin mining—demonstrated in our analysis that even under different market structures, they are providing load flexibility to varying degrees and that there are opportunities for performance optimization. Moreover, this load flexibility is due to the nature of the Bitcoin network's design—a marked departure from the traditional data center business model that deserves more attention in the literature.

We found that when market structures aligned curtailment with marginal emissions, CDCs achieved the greatest emissions reductions. This key finding demonstrates that when deployed strategically, load flexibility can materially reduce carbon intensity on the grid. Policymakers and market designers that link incentive structures to real-time emissions can drive actionable results. Examples include dynamic pricing tied to locational marginal emissions and carbon-responsive demand programs that reward facilities for shifting load away from peak-emissions periods. In regions with carbon-intensive generation, curtailment programs could mitigate the worst environmental impacts without requiring capital-intensive infrastructure changes.

For industry, our results highlight the operational levers available to reduce emissions while maintaining economic viability. Specifically, aligning curtailment schedules with marginal grid emissions and integrating emissions-responsive automation into site-level decision-making can maximize climate benefits. Future strategies should include forecasting tools that balance environmental and Bitcoin-sensitive economic performance in real time.

Amid growing scrutiny of cryptocurrency mining's environmental footprint, this study provides an empirical foundation for rethinking the sector's role in grid decarbonization. This study shows that CDCs are flexible and calls for future energy modeling to carefully integrate these results. Finally, with clearer understanding of their load flexibility potential, constraints, and enabling conditions, this analysis lays the ground for making the most of the unique qualities of CDCs for a new class of climate solutions: ones that are fast, scalable, and rooted in behavioral and market-based shifts. Our work offers foundational analytical metrics and empirical evidence that can inform smarter regulation, targeted investment, and more adaptive decarbonization strategies.

## Declarations

V. P. thanks Casey Martinez and Garrett Morrow for assistance with developing the method for computing avoided emissions for cryptocurrency data centers and for providing access to marginal emission factors. This material is based upon work supported by the Human Rights Foundation, a Bitcoin Policy Institute Bitcoin and Energy Graduate Research Fellowship, and the National Science Foundation under Grant No. 1837021. Any opinions, findings, and conclusions or recommendations expressed in this material are those of the author(s) and do not necessarily reflect the views of the Human Rights Foundation, the Bitcoin Policy Institute, or the National Science Foundation.